\documentclass[12pt]{iopart}
\usepackage{epsf}
\usepackage{pstcol,pst-text}
\begin{document}

\title[Baryon masses and nucleon sigma terms]{Baryon masses
and nucleon sigma terms in manifestly Lorentz-invariant baryon chiral
perturbation theory}

\author{B C Lehnhart\dag, J Gegelia\dag\ddag, S Scherer\dag
\footnote[3]{scherer@kph.uni-mainz.de}}

\address{\dag\ Institut f\"ur Kernphysik, Johannes Gutenberg-Universit\"at
Mainz, J J Becher-Weg 45, D-55099 Mainz, Germany}
\address{\ddag\
High
Energy Physics Institute, Tbilisi State University, University
St.~9, 380086 Tbilisi, Georgia}
\address{}
\address{9 July 2004}

\begin{abstract}
   We discuss the masses of the ground state baryon octet and the
nucleon sigma terms in the framework of manifestly Lorentz-invariant baryon
chiral perturbation theory.
   In order to obtain a consistent power counting for renormalized
diagrams the extended on-mass-shell renormalization scheme is
applied.
\end{abstract}


\submitto{\JPG}


\section{\label{introduction}Introduction}
   Chiral perturbation theory (ChPT) for mesons
\cite{Weinberg:1978kz,Gasser:1983yg,Gasser:1984gg}
has been highly successful in describing the strong interactions of the
pseudoscalar meson octet in the low-energy regime
(for a recent review, see \cite{Scherer:2002tk,Gasser:2003cg}).
   The prerequisite for its success has been a consistent power counting
establishing a straightforward connection between the loop expansion
and the chiral expansion in terms of quark masses and small external
four-momenta at a fixed ratio.
   State-of-the-art calculations have reached the next-to-next-to-leading,
i~e, two-loop order accuracy \cite{Bijnens:2003jw}.
   In general, due to the relatively large mass of the strange quark,
the convergence in the three-flavour sector is somewhat slower in comparison with
the two-flavour sector.
   The extension to processes including one ground state baryon in the
initial and final states was performed in \cite{Gasser:1987rb,Krause:xc}.
   It was found that, when applying the modified minimal subtraction
($\widetilde{\rm MS}$) scheme of ChPT \cite{Gasser:1983yg,Gasser:1984gg}, i~e,
the same renormalization condition as in the mesonic sector, higher-loop diagrams
can contribute to terms as low as ${\cal O}(q^2)$ in the chiral expansion, where
$q$ denotes a small external momentum or a Goldstone boson mass.
   Hence, the correspondence between the loop and chiral expansions seemed
to be lost.
   This problem was eluded in the framework of the heavy-baryon
formulation of ChPT \cite{Jenkins:1990jv,Jenkins:1991ts,Bernard:1992qa},
resulting in a power counting analogous to the mesonic sector.
   The price one pays for giving up manifest Lorentz invariance of the
Lagrangian consists of a) an increasing complexity of the effective Lagrangian
due to $1/m$ corrections and b) the fact that not all of the scattering
amplitudes, evaluated perturbatively in the heavy-baryon framework, show the
correct analytical behaviour in the low-energy region.

   Recently, several methods have been devised to restore a consistent
power counting in a manifestly Lorentz-invariant approach
\cite{Tang:1996ca,Ellis:1997kc,Becher:1999he,Gegelia:1999gf,Becher:2001hv,%
Lutz:2001yb,Fuchs:2003qc}.
   Here, we will concentrate on the so-called extended on-mass-shell
(EOMS) renormalization scheme of \cite{Gegelia:1999gf,Fuchs:2003qc} which
provides a simple and consistent power counting for the renormalized diagrams of
manifestly Lorentz-invariant ChPT.
   The method makes use of finite subtractions of dimensionally
regularized diagrams beyond the standard modified minimal subtraction scheme of
ChPT to remove contributions violating the power counting.
   This is achieved by choosing a suitable renormalization of the parameters
of the most general effective Lagrangian.
   So far the new method has been applied in the two-flavour sector to
a calculation of the nucleon mass \cite{Fuchs:2003qc}, the electromagnetic form
factors \cite{Fuchs:2003ir}, the sigma term and the scalar form factor
\cite{Fuchs:2003kq}.
   Moreover, the EOMS approach allows one to consistently include vector mesons
as explicit degrees of freedom \cite{Fuchs:2003sh} and to reformulate the
infrared renormalization of Becher and Leutwyler \cite{Becher:1999he} so that it
may also be applied to multiloop diagrams
\cite{Schindler:2003xv,Schindler:2003je}.
   In this paper we will discuss the application of the EOMS scheme
to a calculation of the baryon octet masses in the framework of
three-flavor ChPT.
   We will also discuss various nucleon sigma terms.

\section{\label{lagrangian} Lagrangian and power counting}
   In this section we will briefly specify those elements of the effective
Lagrangian in the three-flavour sector which are relevant for the discussion of
the baryon masses and the nucleon sigma terms.
   We will not need to consider external fields except for the
(constant) quark masses, because the sigma terms will be derived from the baryon
masses using the Hellmann-Feynman theorem
\cite{Gasser:1983yg,Gasser:1987rb,Hellmann,Feynman}.
   The effective Lagrangian is written as the sum of a purely mesonic Lagrangian
and the baryonic Lagrangian (including the interactions with the
mesons),
\begin{equation}
\label{leff} {\cal L}_{\rm eff}={\cal L}_{\rm M}+{\cal L}_{\rm B}.
\end{equation}
   Both Lagrangians are organized in a derivative and quark mass expansion,
\begin{eqnarray}
{\cal L}_{\rm M}&=&{\cal L}_2+{\cal L}_4+\cdots,\nonumber\\
{\cal L}_{\rm B}&=&{\cal L}^{(1)}+{\cal L}^{(2)}+\cdots,
\end{eqnarray}
where the subscripts (superscripts) in ${\cal L}_{\rm M}$ (${\cal L}_{\rm B}$)
refer to the order in the chiral expansion.
   The EOMS renormalization scheme \cite{Gegelia:1999gf,Fuchs:2003qc}
is constructed such that, after subtraction, a renormalized
Feynman diagram has a certain chiral order $D$ which is determined
by the following power counting: if $q$ stands for small
quantities such as a meson mass, small external four-momenta of a
meson or small external three-momenta of a baryon, then
interaction terms derived from ${\cal L}^{(k)}$ and ${\cal
L}_{2k}$ count as $q^k$ and $q^{2k}$, respectively, baryon and
meson propagators as $q^{-1}$ and $q^{-2}$, respectively, and a
loop integration in $n$ dimensions as $q^n$.
   Here, we work in the framework of ordinary ChPT, where the quark mass term is
counted as ${\cal O}(q^2)$ (for a different organization
procedure, see \cite{Stern:rg}).

   The relevant lowest-order mesonic Lagrangian is given by
\begin{equation}
\label{l2} {\cal L}_2=\frac{F^2_0}{4}\mbox{Tr}(\partial_\mu
U\partial^\mu U^\dagger) +\frac{F^2_0}{4}\mbox{Tr}(\chi U^\dagger+
U\chi^\dagger),
\end{equation}
where $U$ is a unimodular unitary $(3\times 3)$ matrix containing
the eight Goldstone boson fields,
\begin{equation*}
U=\exp\left(i\frac{\Phi}{F_0}\right), \quad
\Phi=\sum_{a=1}^8\phi_a\lambda_a =\left(
\begin{array}{ccc} \pi^{0} + \frac{1}{\sqrt{3}} \eta & \sqrt{2}
\pi^{+} & \sqrt{2}K^+ \\
\sqrt{2} \pi^{-} & -\pi^{0}+ \frac{1}{\sqrt{3}} \eta & \sqrt{2} K^0 \\
\sqrt{2} K^{-} & \sqrt{2} \bar{K}^{0} & - \frac{2}{\sqrt{3}} \eta
\end{array} \right).
\end{equation*}
   In equation (\ref{l2}), $F_0$ denotes the pion-decay constant in the chiral
limit, $F_\pi=F_0[1+{\cal O}(q^2)]=92.4$ MeV, and $\chi=2 B_0 M$ with $M=
\mbox{diag}(\hat{m},\hat{m},m_s)$ the quark mass matrix, where we assumed perfect
isospin symmetry, $m_u=m_d=\hat{m}$.
    The lowest-order expressions for the squared meson masses read
\numparts
\begin{eqnarray}
M_{\pi,2}^2 & = & 2 B_0 \hat{m}, \label{mpi}\\
M_{K,2}^2 & = & B_0(\hat{m} + m_s), \label{mk}\\
M_{\eta,2}^2 & = & \frac{2}{3} B_0(\hat{m} + 2 m_s), \label{meta}
\end{eqnarray}
\endnumparts
where the subscript 2 refers to ${\cal O}(q^2)$ and the low-energy constant $B_0$
is related to the quark condensate $\langle \bar{q} q\rangle_0$ in the chiral
limit \cite{Gasser:1984gg}.
   At lowest order, the masses of equations
(\ref{mpi}) - (\ref{meta}) satisfy the Gell-Mann--Okubo relation
\begin{equation}
\label{gmo} 4 M_{K,2}^2=3M_{\eta,2}^2 +M_{\pi,2}^2.
\end{equation}

   In order to discuss the baryonic Lagrangian, we collect the
octet of the ground state baryons in the traceless $(3\times 3)$ matrix
\begin{eqnarray}
B & = & \sum_{a=1}^{8}\frac{1}{\sqrt{2}}B_{a}\lambda_{a}  = \left(
\begin{array}{ccc}
\frac{1}{\sqrt{2}} \Sigma^{0} + \frac{1}{\sqrt{6}} \Lambda & \Sigma^{+} & p \\
\Sigma^{-} & - \frac{1}{\sqrt{2}} \Sigma^{0}+ \frac{1}{\sqrt{6}} \Lambda & n \\
\Xi^{-} & \Xi^{0} & - \frac{2}{\sqrt{6}} \Lambda
\end{array} \right).
\end{eqnarray}
   The most general single-baryon Lagrangian is bilinear in $B$
and $\bar{B}=B^\dagger \gamma_0$ and involves the quantities $u=\sqrt{U}$,
$u_\mu$, $\Gamma_\mu$ and $\chi_\pm$ (and covariant derivatives thereof) which,
in the absence of external fields, are given by \numparts
\begin{eqnarray}
u_{\mu} & = & i(u^{\dagger}\partial_{\mu}u -
u\partial_{\mu}u^{\dagger}),\\
\Gamma_{\mu} & = & \frac{1}{2}(u^{\dagger}\partial_{\mu}u +
u\partial_{\mu}u^{\dagger}),\\
\chi_\pm&=&u^\dagger\chi u^\dagger \pm u\chi^\dagger u.
\end{eqnarray}
\endnumparts
   In terms of these building blocks the lowest-order Lagrangian
reads
\begin{equation}
{\cal{L}}^{(1)}= \mbox{Tr}[\bar{B}(iD\hspace{-.65em}/\hspace{.1em}-M_0)B]
-\frac{D}{2}\mbox{Tr}(\bar{B} \gamma^{\mu}\gamma_{5} \{u_{\mu},B\})
-\frac{F}{2}\mbox{Tr}(\bar{B} \gamma^{\mu}\gamma_{5} [ u_{\mu},B]), \label{lb1}
\end{equation}
where the covariant derivative of the baryon field is defined as
$D_{\mu}B  = \partial_{\mu}B+[\Gamma_{\mu},B]$.
   Equation (\ref{lb1}) involves three
low-energy constants, the mass of the baryon octet in the chiral limit, $M_0$,
and the constants $D$ and $F$ which may be determined by fitting the
semi-leptonic decays $B \to B'+e^-+\bar{\nu}_e$ \cite{Borasoy:1998pe}:
\begin{displaymath}
D = 0.80,\quad F = 0.50.
\end{displaymath}
   The Lagrangian at order ${\cal O}(q^2)$ has been discussed in
\cite{Krause:xc,Borasoy:1996bx}.
   Here, we will concentrate on the chiral symmetry breaking part
\begin{equation}
{\cal{L}}^{(2, \rm sb)} =  b_0 \mbox{Tr}(\bar{B}B)\mbox{Tr}(\chi_+)+ b_D
\mbox{Tr}(\bar{B}\{\chi_+,B\})+ b_F\mbox{Tr}(\bar{B}[\chi_+,B]) \label{l2sb}
\end{equation}
relevant for a contact contribution to the self energy at ${\cal O}(q^2)$.
   Moreover, we will also use equation (\ref{l2sb}) for estimating some
one-loop contributions at ${\cal O}(q^4)$.
   The ${\cal O}(q^3)$ Lagrangian does not contribute to the baryon masses.
   Finally, from the ${\cal O}(q^4)$ Lagrangian we only need the
terms with two powers of $\chi_+$ \cite{Borasoy:1996bx},
\begin{eqnarray}
\label{l4chichi} {\cal{L}}^{(4)}&=&d_1\mbox{Tr}(\bar{B}[\chi_+,[\chi_+,B]])
+d_2\mbox{Tr}(\bar{B}[\chi_+,\{\chi_+,B\}])\nonumber\\
&&+d_3\mbox{Tr}(\bar{B}\{\chi_+,\{\chi_+,B\}\})
+d_4\mbox{Tr}(\bar{B}\chi_+)\mbox{Tr}(\chi_+B)\nonumber\\
&&+d_5\mbox{Tr}(\bar{B}[\chi_+,B])\mbox{Tr}(\chi_+)
+d_6\mbox{Tr}(\bar{B}\{\chi_+,B\})\mbox{Tr}(\chi_+)\nonumber\\&&
+d_7\mbox{Tr}(\bar{B}B)[\mbox{Tr}(\chi_+)]^2+\cdots,
\end{eqnarray}
where we have eliminated the $d_8$ term of \cite{Borasoy:1996bx} using a trace
relation \cite{Fearing:1994ga}.

\section{Calculation of the self-energy}
   In order to discuss the baryon masses we need to calculate
the corresponding self-energies $\Sigma_B(p\hspace{-.45em}/)$ which are given as
the one-particle-irreducible perturbative contribution to the two-point functions
\begin{displaymath}
S_B(p)=\frac{1}{p\hspace{-.425em}/-M_0-\Sigma_B(p\hspace{-.425em}/)}.
\end{displaymath}
   The physical masses are defined in terms of the pole
at $p\hspace{-.45em}/=M_B$,
\begin{equation}
\label{pole}
M_B-M_0-\Sigma_B(M_B)=0.
\end{equation}
   From a technical point of view, it is convenient to determine
the cartesian components $\Sigma_{ba}(p\hspace{-.45em}/)$ of the self-energy
(tensor), where the first and second indices refer to the final and initial
baryon lines, respectively.
   Under $V\in \mbox{SU(3)}$ the self-energy transforms as
\begin{displaymath}
\Sigma\mapsto D(V) \Sigma D^{-1}(V),
\end{displaymath}
where $D(V)$ is a real orthogonal representation matrix  with entries
$D_{cd}(V)=\frac{1}{2}\mbox{Tr}(\lambda_c V\lambda_d V^\dagger)$.
   Strangeness conservation  and isospin symmetry imply that the self-energies of the physical
particles are given in terms of the linear combinations \numparts
\begin{eqnarray}
\Sigma_N&=&\Sigma_{44}-i\Sigma_{54},\\
\Sigma_{\Sigma}&=&\Sigma_{33},\\
\Sigma_{\Lambda}&=&\Sigma_{88},\\
\Sigma_{\Xi}&=&\Sigma_{44}+i\Sigma_{54}.
\end{eqnarray}
\label{Sigmadef}
\endnumparts

\subsection{\label{O3calculation} ${\cal O}(q^3)$ calculation}
\begin{figure}
\begin{center}
\epsfbox{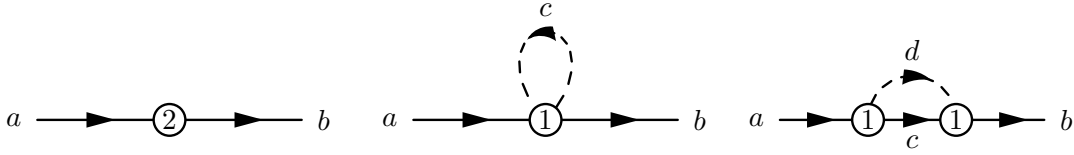}
\end{center}
\caption{\label{dia_O_3:fig} Contributions to the baryon self-energy
up to and including ${\cal O}(q^3)$. The numbers in the interaction blobs denote
the order of the Lagrangian from which they are obtained.}
\end{figure}

   According to the power counting, up to and including ${\cal
O}(q^3)$ the self-energy receives contact contributions from ${\cal L}^{(2)}$
and, in principle, also from ${\cal L}^{(3)}$ as well as one-loop contributions
with vertices from ${\cal L}^{(1)}$.
   The relevant interaction Lagrangians are given by
\numparts
\begin{eqnarray}
{\cal{L}}^{(2,\rm sb)}_{0\phi}& = & 4 B_0 \bar{B}_b B_a \left[ (2\hat{m}+m_s)
\left(b_0 +\frac{2}{3}b_D\right)\delta_{ab}\right.\nonumber\\
&&\left. + \frac{\hat{m}-m_s}{\sqrt{3}}2(d_{ab8}b_D +if_{ab8}
b_F)\right],\label{l2sb0phi}\\
{\cal L}^{(1)}_{1\phi} & = & \frac{1}{F_0} (d_{abc} D + i f_{abc}F) \bar{B}_b
\gamma^{\mu} \gamma_5  B_a  \partial_{\mu} \phi_c,\\
{\cal L}^{(1)}_{2\phi} & = & - \frac{i}{2 F_0^2} f_{abe}
f_{ecd}\bar{B}_b \gamma^{\mu} B_a \phi_c
\partial_{\mu}\phi_d,
\end{eqnarray}
\endnumparts
where $f_{abc}$ and $d_{abc}$ are the structure constants and $d$
symbols of SU(3).
   In equation (\ref{l2sb0phi}) we have separated the Lagrangian into the parts
transforming as an SU(3) singlet and the eighth component of an SU(3) octet,
respectively.
   The Lagrangian ${\cal L}^{(3)}$ does not generate a contact contribution
to the self-energy.
   The subscripts $k\phi$ indicate the number $k$ of Goldstone boson
fields in the interaction terms.

   From the contact contribution we obtain
\begin{equation}
\label{sigmacontact2} \eqalign{ \Sigma^{\rm contact}_{ii}& =  -4B_{0} \left [ b_0
(2\hat{m}+m_{s}) + 2 b_D \hat{m} \right
]\,\,\mbox{for}\,\, i=1,2,3,\\
\Sigma^{\rm contact}_{jj}& =  -4B_{0} \left [ b_0 (2\hat{m}+m_{s}) + b_D
(\hat{m}+m_{s}) \right ]\,\,\mbox{for}\,\,
j=4,5,6,7,\\
\Sigma^{\rm contact}_{54} & =  - \Sigma^{\rm contact}_{45} = \Sigma^{\rm
contact}_{76} = - \Sigma^{\rm contact}_{67}
 =  \mbox{} -4 i B_{0} b_F (\hat{m}-m_{s}), \\
\Sigma^{\rm contact}_{88} & =  -4B_{0} \left [ b_0 (2\hat{m}+m_{s}) + b_D
\frac{2(\hat{m}+2m_{s})}{3} \right ],}
\end{equation}
where the remaining $\Sigma^{\rm contact}_{kl}$ vanish.
   The second type of diagrams in figure \ref{dia_O_3:fig} does not contribute, because
the internal meson lines lead to the same value of indices in the $f$ symbols and
thus vanish.
   Finally, the third contribution of figure \ref{dia_O_3:fig} can be written as
\begin{equation}
\Sigma^{\rm loop}_{ba}(p\hspace{-.425em}/)  =  {\cal F}_{abd} \, {\cal
S}_d(p\hspace{-.425em}/),
\end{equation}
where
\begin{equation}
{\cal F}_{abd}  =  - \sum_{c=1}^8 ( D d_{acd}+i F f_{acd} ) ( D d_{cdb}-i F
f_{cdb} )
\end{equation}
and
\begin{equation}
{\cal S}_d(p\hspace{-.425em}/)  =  - \frac{i}{F_0^2} \int \frac{d^4k}{(2 \pi)^4}
k\hspace{-.45em}/ \gamma_5 \frac{1}{p\hspace{-.425em}/ - k\hspace{-.45em}/ - M_0
+ i0^+} k\hspace{-.425em}/ \gamma_5 \frac{1}{k^2 - M_d^2 + i0^+}.
\end{equation}
   Here,
\begin{equation}
\label{Md2} M_d^2= \left\{
\begin{array}{lll}
M_{\pi,2}^2 & \mbox{for} & d=1,2,3\\
M_{K,2}^2 & \mbox{for} & d=4,5,6,7\\
M_{\eta,2}^2 & \mbox{for} & d=8\\
\end{array}\right.
\end{equation}
   Using the algebra of the gamma matrices and applying dimensional  regularization,
${\cal S}_d(p\hspace{-.425em}/)$ can be expressed in terms of the scalar
integrals of \ref{integrals} as \cite{Fuchs:2003qc}
\begin{eqnarray} {\cal S}_d(p\hspace{-.425em}/) &= &
\frac{1}{F^{2}_{0}} \{
(p\hspace{-.425em}/ + M_0)I_B+(p\hspace{-.425em}/+M_{0})M_d^2 I_{BM_d} \nonumber \\
& & \mbox{} - (p^{2} - M_{0}^{2}) \frac{p\hspace{-.425em}/}{2p^{2}} [I_B -
I_{M_d} + (p^{2} - M_{0}^{2} + M_d^2) I_{BM_d}]\}.
\end{eqnarray}
   According to the power counting, the renormalized diagram, i~e, the sum of
the unrenormalized value of the basic graph and the sum of the counterterm
graphs, should be of ${\cal O}(q^3)$.
   As in \cite{Fuchs:2003qc}, we first perform the modified minimal
subtraction after which the renormalized result is of ${\cal O}(q^2)$ and then an
additional finite subtraction so that the renormalized integral is finally of
${\cal O}(q^3)$:
\begin{eqnarray}
{\cal S}_{d}^{R}(p\hspace{-.425em}/) & = & \frac{1}{F^{2}_{0}} \left[
(p\hspace{-.425em}/-M_{0}) M_d^{2} I_{BM_d}^r + 2M_{0} M_{d}^{2}
I_{BM_d}^R\right.\nonumber\\
&& + (p^{2}-M_{0}^{2}) \frac{p\hspace{-.425em}/}{2p^{2}} I_{M_d}^r -
(p^{2}-M_{0}^{2})^2 \frac{p\hspace{-.425em}/-M_0}{2p^{2}} I_{BM_d}^r \nonumber\\
&&\left. - (p\hspace{-.425em}/-M_{0}^{2})^2 \frac{M_0}{2p^{2}} I_{BM_d}^R -
(p^{2}-M_{0}^{2}) \frac{p\hspace{-.425em}/}{2p^{2}} M_{d}^{2} I_{BM_d}^r \right],
\end{eqnarray}
where the superscripts $r$ and $R$ refer to the $\widetilde{\rm MS}$ and EOMS
subtractions, respectively.
   Defining
\begin{eqnarray*}
{\cal S}_\pi&=& {\cal S}_d\quad{\rm for}\quad d=1,2,3,\\
{\cal S}_K&=& {\cal S}_d\quad{\rm for}\quad d=4,5,6,7,\\
{\cal S}_\eta&=& {\cal S}_d\quad{\rm for}\quad d=8,
\end{eqnarray*}
it is convenient to express the renormalized loop contribution for the physical
baryons as
\begin{equation}
\label{renlooppb} \Sigma_B^{{\rm loop},R}(p\hspace{-.425em}/)={\cal F}_{BM}{\cal
S}^R_M(p\hspace{-.425em}/),
\end{equation}
where $B=N,\Sigma,\Lambda,\Xi$ and a summation over $M=\pi,K,\eta$ is implied.
   The coefficients ${\cal F}_{BM}$ are given in \ref{coefficients}.
   Using
\begin{equation}
\label{SRMO3} {\cal S}^R_M(p\hspace{-.425em}/=M_B)=\frac{M_M^3}{8\pi F_0^2}+{\cal
O}(q^4)
\end{equation}
and inserting the results of equations (\ref{sigmacontact2}) and
(\ref{renlooppb}) into the defining equation (\ref{pole}) we obtain for the
baryon masses at ${\cal O}(q^3)$:
\numparts
\begin{eqnarray}
\label{mn3} M_{N,3} & = & M_0 - 2 (2 M_{K,2}^2 + M_{\pi,2}^2)b_0-4M_{K,2}^2 b_D
+4(M_{K,2}^2-M_{\pi,2}^2)b_F \nonumber \\
& & \mbox{} +\frac{1}{F_0^2} \left[ {\cal{F}}_{N K} \frac{M_{K,2}^3}{8 \pi}+
{\cal{F}}_{N \pi} \frac{M_{\pi,2}^3}{8 \pi} +
{\cal{F}}_{N \eta} \frac{M_{\eta,2}^3}{8 \pi} \right], \\
\label{msigma3} M_{\Sigma,3} & = &M_0 - 2 (2 M_{K,2}^2 +
M_{\pi,2}^2)b_0 -4M_{\pi,2}^2 b_D\nonumber\\
& & \mbox{} +\frac{1}{F_0^2} \left[ {\cal{F}}_{\Sigma K} \frac{M_{K,2}^3}{8 \pi}
+ {\cal{F}}_{\Sigma \pi} \frac{M_{\pi,2}^3}{8
\pi} +{\cal{F}}_{\Sigma \eta} \frac{M_{\eta,2}^3}{8 \pi} \right], \\
\label{mlambda3}
 M_{\Lambda,3} & = &M_0 - 2 (2 M_{K,2}^2 + M_{\pi,2}^2)b_0
 -4 M_{\eta,2}^2 b_D\nonumber\\
& & \mbox{} +\frac{1}{F_0^2} \left[ {\cal{F}}_{\Lambda K} \frac{M_{K,2}^3}{8
\pi}+ {\cal{F}}_{\Lambda \pi} \frac{M_{\pi,2}^3}{8 \pi} + {\cal{F}}_{\Lambda
\eta} \frac{M_{\eta,2}^3}{8 \pi} \right],\\
\label{mxi3}
 M_{\Xi,3} & = &M_0 - 2 (2 M_{K,2}^2 + M_{\pi,2}^2)b_0
 -4M_{K,2}^2 b_D -4(M_{K,2}^2-M_{\pi,2}^2)b_F \nonumber \\
& & \mbox{} + \frac{1}{F_0^2} \left[ {\cal{F}}_{\Xi K} \frac{M_{K,2}^3}{8 \pi}+
{\cal{F}}_{\Xi \pi} \frac{M_{\pi,2}^3}{8 \pi} + {\cal{F}}_{\Xi \eta}
\frac{M_{\eta,2}^3}{8 \pi} \right].
\end{eqnarray}
\endnumparts
   We made use of equations (\ref{mpi}) and (\ref{mk}) to express the quark
   mass terms of the contact contribution in terms of the lowest-order Goldstone
   boson masses.
   Even without performing a numerical analysis, a few properties of equations
(\ref{mn3}) - (\ref{mxi3}) are easily shown:
   For equal quark masses $\hat m=m_s$ the baryon octet becomes degenerate.
   Moreover, in the chiral limit, its mass reduces to $M_0$.
   The $b_0$ term (and part of the $b_D$ term) results in a mass shift of the complete
baryon octet.
   To disentangle the parameters $M_0$ and $b_0$ one
needs additional input such as the pion-nucleon sigma term to be discussed below.
   If one considers the predictions up to and including ${\cal O}(q^2)$
   only, the results satisfy the Gell-Mann--Okubo mass formula
   \begin{equation}
2(M_{N,2} + M_{\Xi,2})=3 M_{\Lambda,2}+M_{\Sigma,2}.
\end{equation}
   This is no longer true, once the ${\cal O}(q^3)$ contribution is included,
   because the Gell-Mann--Okubo mass formula is derived from first-order
   perturbation theory in the SU(3) symmetry breaking quark mass
   term proportional to $\lambda_8$.\footnote{Using the empirical values
   of footnote +, the deviation from the Gell-Mann--Okubo mass formula
   is very small
   \begin{displaymath}
\Delta_{\rm GMO}\equiv\frac{2(M_N+M_\Xi)}{3M_\Lambda+M_\Sigma}-1=-0.6\,\%.
\end{displaymath}}

\subsection{Sigma terms}
   Sigma terms provide a sensitive measure of explicit chiral symmetry breaking in
QCD because "they are corrections to a null result in the chiral limit rather
than small corrections to a non-trivial result" \cite{Pagels:se} (see
\cite{Reya:gk} and \cite{Gensini:1998au} for an early and a recent review,
respectively).
   The so-called sigma commutator is defined as
\begin{equation}
\label{defSig} \sigma^{ab}(x)\equiv [Q^a_A(x_0),[Q^b_A(x_0),{\cal H}_{\rm sb}
(x)]],
\end{equation}
where $Q_A^c=Q_R^c-Q_L^c$ denotes one of the eight axial charge operators (see,
e~g, \cite{Scherer:2002tk} for further details) and
\begin{displaymath}
{\cal H}_{\rm sb}=\bar{q}M q=\hat{m}(\bar{u}u+ \bar{d}d) + m_s \bar{s}s
\end{displaymath}
is the chiral symmetry breaking mass term of the QCD Hamilton density in the
isospin symmetrical limit.
   Using equal-time (anti-) commutation relations, equation (\ref{defSig})
can be written as \cite{Lyubovitskij:2000sf}
\begin{displaymath}
\sigma^{ab}(x)=\bar{q}(x)\{\frac{\lambda^a}{2},\{\frac{\lambda^b}{2},M\}\}q(x)
\end{displaymath}
yielding for the flavor-diagonal pieces
\numparts
\begin{eqnarray}
\sigma^{11}=\sigma^{22}=\sigma^{33}=\hat{m}(\bar{u}u+\bar{d}d),\\
\sigma^{44}=\sigma^{55}=\frac{\hat{m}+m_s}{2}(\bar{u}u+\bar{s}s),\\
\sigma^{66}=\sigma^{77}=\frac{\hat{m}+m_s}{2}(\bar{d}d+\bar{s}s),\\
\sigma^{88}=\frac{1}{3}[\hat{m}(\bar{u}u+\bar{d}d)+4m_s\bar{s}s],\\
\sigma^{38}=\sigma^{83}=0.
\end{eqnarray}
\endnumparts
   Here, we will be concerned with the nucleon sigma terms
defined in terms of proton matrix elements, \numparts
\begin{eqnarray}
\sigma_{\pi N} \equiv \frac{1}{2M_p}\langle p |\sigma^{11}(0)|p\rangle, \\
\sigma_{K N}^u \equiv \frac{1}{2M_p}\langle p |\sigma^{44}(0)|p\rangle, \\
\sigma_{K N}^d \equiv \frac{1}{2M_p}\langle p |\sigma^{66}(0)|p\rangle, \\
\sigma_{\eta N} \equiv \frac{1}{2M_p}\langle p |\sigma^{88}(0)|p\rangle,
\end{eqnarray}
\endnumparts
where $M_p$ is the proton mass.\footnote{The factor $2M_p$ in the denominator
results from our normalization of the states and of the Dirac spinors:
\begin{displaymath}
\langle \vec{p}\,',s'|\vec{p},s\rangle=2
E(\vec{p}\,)(2\pi)^3\delta^3(\vec{p}\,'-\vec{p}\,)\delta_{s's},\quad
\bar{u}(\vec{p},s')u(\vec{p},s)=2M_p\delta_{s's}.
\end{displaymath}}
   Instead of the flavour components $\sigma_{K N}^u$ and $\sigma_{K N}^d$ one
often also discusses the isoscalar and isovector combinations
\numparts
\begin{eqnarray}
\sigma_{K N}^{I=0}&=&\frac{1}{2}(\sigma_{K N}^u+\sigma_{K N}^d)=
\frac{\hat{m}+m_s}{8M_p}\langle p|(\bar{u}u+\bar d d+2 \bar ss)|p\rangle,
\label{isoscalar}\\
\sigma_{K N}^{I=1}&=&\frac{1}{2}(\sigma_{K N}^u-\sigma_{K N}^d)=
\frac{\hat{m}+m_s}{8M_p}\langle p|(\bar{u}u-\bar d d)|p\rangle.
\label{isovector}
\end{eqnarray}
\endnumparts
   Using the first-order mass formula of \cite{Gasser:1982ap},
\begin{displaymath}
M^2_\Xi-M^2_\Sigma\approx (m_s-\hat{m})\langle p|(\bar{u}u-\bar dd)|p\rangle,
\end{displaymath}
the isovector kaon-nucleon sigma term is expected to be small
\cite{Gasser:2000wv},\footnote{Numerical estimates are performed with $M_N=939$
MeV, $M_\Sigma=1193$ MeV, $M_\Lambda=1116$ MeV, $M_\Xi=1318$ MeV, $M_\pi=137$
MeV, $M_K=496$ MeV and $M_\eta=567$ MeV as obtained from equation (\ref{gmo}).}
\begin{equation}
\label{isovectorestimate} \sigma_{KN}^{I=1}\approx \frac{r+1}{r-1}
\frac{M^2_\Xi-M^2_\Sigma}{8M_p}\approx 45\, \mbox{MeV},
\end{equation}
where $r\equiv m_s/\hat m\approx 25$.
   The pion-nucleon sigma term and the strangeness matrix element
$S\equiv (m_s/2M_p)\langle p|\bar{s}s|p\rangle$ are obtained from the proton mass
by using the Hellmann-Feynman theorem\footnote{Given a Hermitian operator
$H(\lambda)$ depending on some parameter(s) $\lambda$ and a normalized eigenstate
$|\alpha(\lambda)\rangle$ with eigenvalue $E(\lambda)$ the Hellmann-Feynman
theorem \cite{Hellmann,Feynman} states that
\begin{displaymath}
\frac{\partial E(\lambda)}{\partial \lambda}=\langle\alpha(\lambda)|
\frac{\partial H(\lambda)}{\partial \lambda}|\alpha(\lambda)\rangle.
\end{displaymath}}
    for $H_{\rm QCD}$:
\numparts
\begin{eqnarray}
\sigma_{\pi N}&=&\hat{m}\frac{\partial M_N}{\partial \hat{m}},\\
S&=&m_s\frac{\partial M_N}{\partial m_s}.
\end{eqnarray}
\endnumparts
   At {\em leading} order in the quark mass expansion (${\cal O}(q^2)$),
$\sigma_{\pi N}$ and $S$ can be interpreted as the contribution to the nucleon
mass which is due to the masses of up and down quarks and of the strange quark,
respectively.
   However, such an interpretation is no longer true beyond leading order.
   Using the results of equation (\ref{mn3}) we obtain at ${\cal O}(q^3)$
\numparts
\begin{eqnarray}
\label{sigmapiN3} \sigma_{\pi N,3}&=&-2M_{\pi,2}^2(2b_0+b_D+b_F) \nonumber\\&&
+\frac{M_{\pi,2}^2}{16\pi F_0^2}\left[\frac{3}{2}M_{K,2}{\cal F}_{NK}
+3M_{\pi,2}{\cal F}_{N\pi}+M_{\eta,2}{\cal F}_{N\eta}\right],\\
\label{S3} S_3&=&\left.\left(M_{K,2}^2-\frac{1}{2}M_{\pi,2}^2\right)
\right[-4(b_0+b_D-b_F)\nonumber\\
&& \left.+\frac{1}{8\pi F_0^2}\left(\frac{3}{2}M_{K,2}{\cal
F}_{NK}+2M_{\eta,2}{\cal F}_{N\eta}\right)\right].
\end{eqnarray}
\endnumparts
The so-called strangeness content or strangeness fraction of the proton is
defined as \cite{Gasser:1980sb}
\begin{equation}
\label{scy} y\equiv\frac{2\langle p|\bar s s|p\rangle}{\langle p|(\bar u u+\bar d
d)|p\rangle} =\frac{2}{r}\frac{S}{\sigma_{\pi N}} =2 \frac{ \frac{\partial
M_N}{\partial m_s}}{\frac{\partial M_N}{\partial \hat m}},
\end{equation}
where, again, we made use of the Hellmann-Feynman theorem.
   Finally, the isoscalar kaon-nucleon sigma term and the eta-nucleon sigma
term can then be re-expressed as \numparts
\begin{eqnarray}
\sigma_{KN}^{I=0}&=&\frac{1}{4}(1+r)(1+y)\sigma_{\pi N},\\
\sigma_{\eta N}&=&\frac{1}{3}(1+2ry)\sigma_{\pi N}.
\end{eqnarray}
\endnumparts

\subsection{Estimate of ${\cal O}(q^4)$ contributions}
\begin{figure}
\begin{center}
\epsfbox{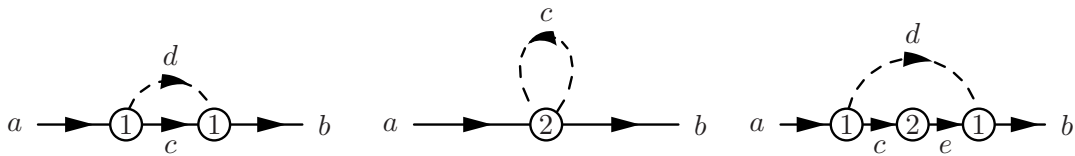}
\end{center}
\caption{\label{dia_O_4:fig} One-loop contributions to the baryon self-energy
at ${\cal O}(q^4)$. The numbers in the interaction blobs denote the order of the
Lagrangian from which they are obtained.}
\end{figure}

   A complete analysis of the baryon masses and the nucleon sigma terms
at ${\cal O}(q^4)$ is beyond the scope of the present work.
   It would require knowledge of additional low-energy constants of the
${\cal O}(q^2)$ Lagrangian beyond the constants $b_0$, $b_D$ and $b_F$ of
equation (\ref{l2sb}) for which we do not have sufficient information.
   Here we will concentrate on an estimate of the one-loop contributions shown
in figure \ref{dia_O_4:fig}.
    To that end we need to identify the ${\cal O}(q^4)$ part of equation
(\ref{SRMO3}) (first diagram of figure \ref{dia_O_4:fig})
\begin{equation}
\label{SRMO4} {\cal S}^R_{M(4)}(p\hspace{-.425em}/=M_B)=\frac{M_M^4}{8\pi^2 F_0^2
M_0}\left[\ln\left(\frac{M_M}{M_0}\right)-1\right].
\end{equation}
   From the second diagram of figure \ref{dia_O_4:fig} we only consider those
contributions which originate from the chiral symmetry breaking vertices of
equation (\ref{l2sb}).
   The contributions of the last diagram of figure \ref{dia_O_4:fig} are obtained
by a direct calculation using the ${\cal O}(q^2)$ contact results of equation
(\ref{sigmacontact2}).\footnote{Alternatively, one may also use the well-known
method of shifting $M_0\delta_{ij}\to M_0\delta_{ij}+\Sigma^{\rm contact}_{ij}$
in the third contribution of figure \ref{dia_O_3:fig} to identify the relevant
terms \cite{Becher:1999he}.}
   Finally, a calculation at ${\cal O}(q^4)$ receives a contact contribution
originating from the ${\cal O}(q^4)$ Lagrangian of equation (\ref{l4chichi}).
   The contact contributions of the $d_5$, $d_6$ and $d_7$ terms are obtained
by the replacements
\begin{equation}
\eqalign{
b_F&\to b_F+4B_0(2\hat m+m_s)d_5,\\
b_D&\to b_D+4B_0(2\hat m+m_s)d_6,\\
b_0&\to b_0+4B_0(2\hat m+m_s)d_7.}
\end{equation}
On the other hand, the $d_1$ - $d_4$ terms result in \numparts
\begin{eqnarray}
M_N^{(4)\rm contact}&=&-16 B_0^2[(m_s-\hat m)^2 d_1-(m_s^2-\hat m^2)d_2+(m_s+\hat m)^2 d_3],
\label{mn4c}\\
M_{\Sigma}^{(4)\rm contact}&=&-16B_0^2 4 \hat m^2 d_3,\label{ms4c}\\
M_{\Lambda}^{(4)\rm contact}&=&-16B_0^2
[\frac{4}{3}(\hat m^2+2m_s^2)d_3+\frac{2}{3}(m_s-\hat m)^2 d_4],\label{ml4c}\\
M_{\Xi}^{(4)\rm contact}&=&-16B_0^2[(m_s-\hat m)^2d_1+(m_s^2-\hat
m^2)d_2+(m_s+\hat m)^2d_3].\label{mx4c}
\end{eqnarray}
\endnumparts
   In the SU(3) limit $m=\hat m=m_s$, equations (\ref{mn4c}) - (\ref{mx4c})
generate a common contribution $M^{4,\rm contact}=-64 B_0^2 m^2 d_3$.
   For a complete chiral expansion of the baryon masses to second order in the
quark masses to be used in unquenched three-flavour lattice simulations we refer
to the recent work by Frink and Mei{\ss}ner \cite{Frink:2004ic}.

\section{Results and discussion}
   In order to determine the parameters $M_0$, $b_0$, $b_D$ and $b_F$ we consider
equations (\ref{mn3}) - (\ref{mxi3}) in combination with equation
(\ref{sigmapiN3}) which allows one to unravel $M_0$ and $b_0$.
   The set of equations can be written as
 \begin{equation}
 \label{setofequationsO3}
\left( \begin{array}{cccc} 1 & a_{b_0} & -4M_{K,2}^2 &
a_{b_F}\\
1 &
a_{b_0} & -4 M_{\pi,2}^2 & 0 \\
1 &
a_{b_0} & -4 M_{\eta,2}^2 & 0 \\
1 & a_{b_0} & -4M_{K,2}^2 &
-a_{b_F}\\
0 &-4 M_{\pi,2}^2 & -2 M_{\pi,2}^2 & -2 M_{\pi,2}^2\\
\end{array} \right) \left( \begin{array}{c} M_0 \\
b_0\\ b_D \\ b_F\\ \end{array} \right)=
\left( \begin{array}{c} \Delta M_{N} \\
\Delta M_{\Sigma}\\
\Delta M_{\Lambda}\\
\Delta M_{\Xi}\\
\Delta \sigma_{\pi N}
\end{array} \right),
\end{equation}
where $a_{b_0}=-2(2M_{K,2}^2+M_{\pi,2}^2)$, $a_{b_F}=4(M_{K,2}^2-M_{\pi,2}^2)$
and $\Delta M_N=M_{N,3}-M_{N(3)}$ is the difference between the nucleon mass up
to and including ${\cal O}(q^3)$ and its ${\cal O}(q^3)$ contribution.
   Note that the ${\cal O}(q^3)$ contribution is fixed in terms of the parameters
$D$, $F$, $F_0$ and the lowest-order Goldstone boson masses.
   The remaining $\Delta$ quantities are defined analogously.

   Rather than selecting three of the mass equations we perform a least squares
fit to all equations.
   For the masses at ${\cal O}(q^3)$ we insert the physical values (see footnote
   +).
   The resulting parameters are given in table \ref{tablefit} for three different
analyses of the (empirical) value of the pion-nucleon sigma term
\cite{Gasser:1990ce,Buettiker:1999ap,Pavan:2001wz} as well as a different value
of $F_0$ in order to compare with \cite{Ellis:1999jt}.
   For fixed parameters $D$, $F$ and $F_0$, the values of $b_D$ and $b_F$ always
come out the same while the results for $M_0$ and $b_0$ are linear functions of
the pion-nucleon sigma term,
\begin{equation}
\label{M0b0}
 M_0=M-\frac{r+2}{2}\sigma_{\pi N},\quad b_0=b-\frac{1}{4
M^2_{\pi,2}}\sigma_{\pi N},
\end{equation}
where $r=m_s/\hat m\approx 25$ and $M=1651$ MeV and $b=-0.235$ GeV$^{-1}$ for
$(D,F,F_0)=(0.80,0.50,92.4\,\mbox{MeV})$.
   Due to the large nonanalytic contributions at ${\cal O}(q^3)$
the parameters $b_0$, $b_D$ and $b_F$ are rather different than in an ${\cal
O}(q^2)$ determination (see, e g, \cite{Thomas:kw} and equations (\ref{mn3}) -
(\ref{mxi3}) and (\ref{sigmapiN3}) for $\sigma_{\pi N}=45$ MeV)
\begin{eqnarray*}
M_{\Sigma,2}-M_{\Lambda,2}=\frac{16}{3}(M^2_{K,2}-M^2_{\pi,2})b_D &\Rightarrow&
b_D=0.064\,\mbox{GeV}^{-1},\\
M_{N,2}-M_{\Xi,2}=8(M^2_{K,2}-M^2_{\pi,2})b_F&\Rightarrow&
b_F=-0.21\,\mbox{GeV}^{-1},\\
\sigma_{\pi N,2}=-2M_{\pi,2}^2(2b_0+b_d+b_F)&\Rightarrow&
b_0=-0.53\,\mbox{GeV}^{-1}.
\end{eqnarray*}
   Here, the effect is even larger than in the SU(2) sector where
differences by factors of about 1.5 were generally observed for the determination
of the low-energy constants at ${\cal O}(q^2)$ and to one-loop accuracy ${\cal
O}(q^3)$ \cite{Bernard:1995dp,Bernard:1997gq}.
\begin{table}
\caption{\label{tablefit} Mass $M_0$ of the baryon octet in the chiral limit
and low-energy constants $b_0$, $b_D$ and $b_F$ as obtained from a $\chi^2$ fit
to equation (\ref{setofequationsO3}). We use three different values for the sigma
term as input. For comparison we also show the results using the parameters of
\cite{Ellis:1999jt}. Note that the differences in the output values of the masses
are due to round-off errors in $M_0$, $b_0$, $b_D$ and $b_F$.} \lineup
\begin{tabular}[t]{lllll}
 \br ($\sigma_{\pi N},F_0)$ [MeV] & (40 \cite{Buettiker:1999ap}, 92.4) &
 (45 \cite{Gasser:1990ce}, 92.4) & (64 \cite{Pavan:2001wz}, 92.4) & (45
\cite{Gasser:1990ce}, 103 \cite{Ellis:1999jt})\\
$M_0$ [MeV] &1107& 1039 & 781 & 965 \\
$b_0$ [GeV$^{-1}$] & $-0.767$ & $-0.833$ & $-1.09$ & $-0.774$ \\
$b_D$ [GeV$^{-1}$] & $0.0314$ & $0.0314$ & $0.0314$ & $0.0381$ \\
$b_F$ [GeV$^{-1}$] &$-0.638$ & $-0.638$ &$-0.638$ &$-0.554$ \\
$M_N$ [MeV] & 941 & 940 & 945 & 942 \\
$M_\Sigma$ [MeV] & 1192 & 1192 & 1196 & 1193 \\
$M_\Lambda$ [MeV] & 1113 & 1113 & 1117 & 1114 \\
$M_\Xi$ [MeV] & 1320 &1319 & 1324 & 1321\\
$\sigma_{\pi N}$ [MeV] & $40.0$ & $45.0$ & $64.3$ & $45.0$ \\
\br
\end{tabular}
\end{table}
   The results for the strangeness matrix element $S$, the strangeness content
$y$ and the various nucleon sigma terms at ${\cal O}(q^3)$ are summarized in
table \ref{sigmaterms}.
   These results are extremely sensitive to which value of the pion-nucleon
sigma term is used as an input.
\begin{table}
\caption{\label{sigmaterms} Strangeness matrix element $S$, strangeness content
$y$ and nucleon sigma terms at ${\cal O}(q^3)$.} \lineup
\begin{tabular}{lllll}
\br ($\sigma_{\pi N},F_0)$ [MeV] & (40 \cite{Buettiker:1999ap}, 92.4) & (45
\cite{Gasser:1990ce}, 92.4) & (64 \cite{Pavan:2001wz}, 92.4) & (45
\cite{Gasser:1990ce}, 103 \cite{Ellis:1999jt})\\
$S$ [MeV] & $-376$ & $-313$ & $-70.0$ & $-205$\\
$y$ & $-0.751$ & $-0.557$ & $-0.0872$ & $-0.364$\\
$\sigma_{KN}^{I=1}$ [MeV] &45&45&45&45\\
$\sigma_{KN}^{u}$ [MeV] &111&176&427&232\\
$\sigma_{KN}^{d}$ [MeV] &21&86&337&142\\
$\sigma_{\eta N}$ [MeV] &$-483$&$-399$&$-71.1$&$-255$ \\
\br
\end{tabular}
\end{table}

   As an example let us discuss the various contributions to the
pion-nucleon sigma term of equation (\ref{sigmapiN3}) using the second set of
parameters in table \ref{tablefit}:
\begin{equation}
\label{sigmapiN3num} \sigma_{\pi N,3}= (85.3 - 22.8 - 16.5 -1.0)\,\mbox{MeV},
\end{equation}
where the first term is the ${\cal O}(q^2)$ contribution and the remaining terms
result from pion, kaon and eta contributions at ${\cal O}(q^3)$.
   It is instructive to compare equation (\ref{sigmapiN3num}) with the corresponding
result of an SU(2) calculation at ${\cal O}(q^3)$
\cite{Bernard:1992qa,Becher:2001hv,Fuchs:2003kq}\footnote{The constants $c_1$ of
\cite{Bernard:1992qa,Becher:2001hv,Fuchs:2003kq} refer to the corresponding
renormalization schemes of HBChPT, infrared and EOMS renormalization,
respectively. The functional form is different for the modified minimal
subtraction ($\widetilde{\rm MS}$) scheme applied in \cite{Gasser:1987rb} (see
their equations (6.5) and (7.5)).}
\begin{displaymath} \sigma_{\pi N}=-4 c_1
M^2-\frac{9 \stackrel{\circ}{g}_A^2}{64\pi F^2}M^3+\cdots,
\end{displaymath}
where $M^2=2B\hat{m}$ is the lowest-order expression for the squared pion mass,
$F$ and $\stackrel{\circ}{g}_A$ refer to the $\mbox{SU}(2)_L\times\mbox{SU}(2)_R$
chiral limit of the pion-decay constant and the axial-vector coupling constant,
respectively.
  In the SU(2) framework, the strange quark mass is considered large and a comparison
with equation (\ref{sigmapiN3}) would yield
\begin{displaymath}
-4 c_1 M^2_\pi=-2M^2_\pi(2b_0+b_D+b_F)+\frac{M_\pi^2}{16\pi
F_\pi^2}\left(\frac{3}{2} M_{K,2}{\cal F}_{NK}+M_{\eta,2}{\cal F}_{N\eta}\right)
\end{displaymath}
which is smaller than the ${\cal O}(q^2)$ contribution in the SU(3) framework.
   This example serves as an illustration for the mechanism shifting contributions
which are of higher order in the SU(3) sector to lower orders in the SU(2)
sector.

   A popular first-order estimate of the pion-nucleon sigma term makes use
of the predictions for the baryon masses at ${\cal O}(q^2)$.
   If, in addition, one assumes a suppression of the strangeness matrix
element in the spirit of the Zweig rule, i~e
$S\approx 0$, the
   pion-nucleon sigma term may be re-written as \cite{Cheng:wm,ChengLi}
\begin{equation}
\sigma_{\pi N}\approx 3 \frac{\hat{m}}{\hat{m}-m_s}\frac{1}{2 M_p}\langle p| c_8
u_8|p\rangle,
\end{equation}
where $c_8 u_8=(\hat{m}-m_s)(\bar{u}u+\bar{d}d-2\bar{s}{s})/3$ is the SU(3)
symmetry breaking part of the mass term.
   At ${\cal O}(q^2)$ the corresponding contribution to the nucleon mass reads
\begin{displaymath}
4(M^2_{K,2}-M^2_{\pi,2})\left(-\frac{b_D}{3}+b_F\right)=M_{\Lambda,2}-M_{\Xi,2},
\end{displaymath}
yielding the estimate \cite{ChengLi}
\begin{equation}
\sigma_{\pi N}\approx 3
\frac{\hat{m}}{\hat{m}-m_s}(M_{\Lambda,2}-M_{\Xi,2})=25.3\,\mbox{MeV}.
\end{equation}
   It was already noted in \cite{Cheng:wm} that such lowest-order estimates have
to be treated with great care and that the application of the Zweig rule may
receive a large correction as is supported in the present case by the range of
possible values of $S$ in table \ref{sigmaterms}.
   Furthermore, an inadequate use of the lowest-order formula
   \begin{displaymath}
   M_N\approx M_0+\sigma_{\pi N}+S
   \end{displaymath}
would lead to the values $(1275,1207,945,1099)$ MeV for $M_0$ which clearly
overestimate the corresponding ``correct'' values $(1107,1039,781,965)$ MeV of
table \ref{tablefit}.

\begin{figure}
\begin{center}
\epsfbox{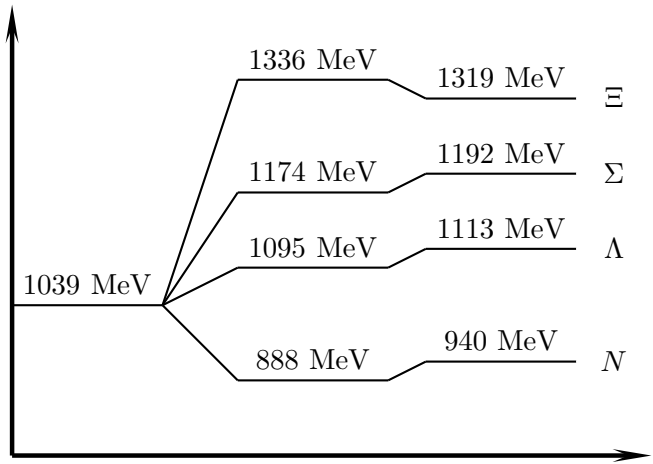}
\end{center}
\caption{\label{massen:fig} Mass level diagram depending on the various
symmetries. Left panel: $\mbox{SU(3)}_L\times\mbox{SU(3)}_R$ symmetry; middle
panel: $\mbox{SU(2)}_L\times\mbox{SU(2)}_R$ symmetry; right panel:
$\mbox{SU(2)}_V$ symmetry.}
\end{figure}

   Using the second set of parameters of table \ref{tablefit},
in figure \ref{massen:fig} we show how ``switching on'' the quark masses affects
the masses of the baryon octet.
   In the chiral limit all masses reduce to $M_0=1039$ MeV.
   Keeping the up and down quarks massless, we still have an
$\mbox{SU(2)}_L\times\mbox{SU(2)}_R$ symmetry resulting in
\begin{equation}
M_{\pi,2}^2 = 0,~~~ M_{K,2}^2 ~=~ B_0 m_s,~~~ M_{\eta,2}^2 ~=~ \frac{4}{3} B_0
m_s ~=~ \frac{4}{3} M_{K,2}^2.
\end{equation}
   The corresponding values of the mass spectrum are shown in the middle panel of
figure \ref{massen:fig} ($M_\pi=0$, $M_K=486$ MeV and $M_\eta=562$ MeV), while
the final results, exhibiting only an $\mbox{SU(2)}_V$ symmetry, are shown in the
right panel.
   Using $\sigma_{\pi N}=45$ MeV as input, an analysis of the nucleon mass
   at ${\cal O}(q^4)$ in the framework of SU(2) chiral perturbation theory
yields an estimate for the nucleon mass in the chiral limit (at fixed $m_s\neq
0$) of $m\approx 883$ MeV \cite{Fuchs:2003kq} which is in surprisingly good
agreement with the 888 MeV of figure  \ref{massen:fig}.
   The pattern of figure \ref{massen:fig} suggests to
treat the explicit symmetry breaking due to the average up and down quark mass
$\hat{m}$ and the strange quark mass $m_s$ on a different footing.
   In this context it might be worthwhile to explore a perturbative series up to
and including terms of second order in the strange quark mass but neglecting
terms of second order in $\hat{m}$.

   Let us now turn to an estimate of (some) ${\cal O}(q^4)$ corrections.
   We stress that our analysis can only be indicative, because we do not fully
analyze the one-loop contribution due to ${\cal L}^{(2)}$ and, correspondingly,
the contact contribution from  ${\cal L}^{(4)}$.
   In table \ref{massesO4} we show the individual contributions to the baryon masses
for the first set of parameters of table~\ref{tablefit}.
   Large cancellations between the contributions $M_{(2)}$ and $M_{(3)}$ at ${\cal
O}(q^2)$ and ${\cal O}(q^3)$, respectively, are a well-known feature
\cite{Borasoy:1996bx,Donoghue:1998bs}.
   On the other hand, the partial ${\cal O}(q^4)$ contributions seem to indicate
a convergence, albeit a slow one.
\begin{table}
\caption{\label{massesO4}
Baryon masses in MeV and their individual contributions. Note that the free
parameters have been fit so that the masses at ${\cal O}(q^3)$ essentially agree
with the physical masses. The ${\cal O}(q^4)$ is incomplete and is to be
understood as an order-of-magnitude estimate.}
\begin{tabular}[t]{llllll}
\br & Chiral limit: $M_0$ & $M_{(2)}$ & $M_{(3)}$ & Sum at ${\cal O}(q^3)$: $M_3$ & $M_{(4)}$ (incomplete)\\
$M_N$&1039&240&$-339$&940&287\\
$M_\Sigma$&1039&849&$-696$&1192&207\\
$M_\Lambda$&1039&811&$-737$&1113&594\\
$M_\Xi$&1039&1400&$-1120$&1319&439\\
$\sigma_{\pi N}$&---&85&$-40$&45&51\\ \br
\end{tabular}
\end{table}
   Finally, we have also used the admittedly incomplete expressions
for the four masses at ${\cal O}(p^4)$ to re-fit the parameters.
   One can then estimate the pion-nucleon sigma term rather than using
it as an input.
   We have fixed $\mu=1$ GeV and obtained $M_0=647$ MeV and $\sigma_{\pi N}=60.4$
MeV.
   Note that this value of $M_0$ is 23 \% smaller than $M_0=836$ MeV which would have
been obtained at ${\cal O}(q^3)$ for the corresponding sigma term (see equation
(\ref{M0b0})).

    Last but not least, we would like to compare our results with other calculations in the
framework of chiral perturbation theory (see table \ref{massesO3oc}). (For an
overview of sigma-term calculations in the framework of other approaches and
various models see, e~g, \cite{Lyubovitskij:2000sf,Schweitzer:2003sb} and
references therein.)
   At ${\cal O}(q^3)$, the EOMS expressions for the masses and the sigma terms are the
same as in the heavy-baryon formulation \cite{Bernard:1993nj} but differ from the
infrared regularization \cite{Ellis:1999jt}, where the one-loop contributions are
generally smaller, because the relevant loop integral (see equation (10) of
\cite{Ellis:1999jt}) is not expanded.
   Similarly, the so-called long-distance regularization of
   \cite{Donoghue:1998rp}, when applied to a calculation of the baryon masses
\cite{Donoghue:1998bs}, generates a faster converging series.
   This corresponds to a re-summation of terms which would appear at higher orders in the
conventional framework.
   On the other hand, in a full ${\cal O}(q^4)$ calculation,
the corresponding expressions will, in general, differ in all schemes
\cite{Borasoy:1996bx,Frink:2004ic,Ellis:1999jt} (see \cite{Fuchs:2003ir} for the
SU(2) sector).
\begin{table}
\caption{\label{massesO3oc}
Baryon masses at third order in units of MeV and their individual contributions
in the framework of long-distance regularization (LDR) \cite{Donoghue:1998rp},
infrared regularization (IR) \cite{Ellis:1999jt} and EOMS.}
\begin{tabular}[t]{llll}
\br
& LDR & IR & EOMS \\
$M_N$ & $1143-237+34=940$ & $733+342-160=915$ & $1039+240-339=940$\\
$M_\Sigma$ & $1143-5+53=1191$ & $733+919-494=1158$ & $1039+849-696=1192$\\
$M_\Lambda$ & $1143-86+57=1114$ & $733+671-201=1204$ & $1039+811-737=1113$\\
$M_\Xi$ & $1143+106+77=1326$ & $733+1124-589=1268$ & $1039+1400-1120=1319$\\
\br
\end{tabular}
\end{table}

\section{Summary and outlook}
   We have discussed the masses of the ground state baryon octet and the
nucleon sigma terms in the framework of manifestly Lorentz-invariant baryon
chiral perturbation theory applying the EOMS renormalization scheme of
\cite{Fuchs:2003ir}.
   At ${\cal O}(q^3)$ the results are identical with those
of the heavy-baryon formulation \cite{Bernard:1993nj}.
   We have performed a least squares fit for the parameters $M_0$,
$b_0$, $b_D$ and $b_F$ using the empirical masses and three different values of
the pion-nucleon sigma term as input (see table \ref{tablefit}).
   We have then discussed the strangeness matrix element $S$, the strangeness
content $y$, the remaining nucleon sigma terms and the baryon octet mass in the
chiral limit as functions of the pion-nucleon sigma term (see table
\ref{sigmaterms}).
   We have also estimated some ${\cal O}(q^4)$ contributions pointing towards
a slow convergence (see table \ref{massesO4}).
   We have stressed that our analysis at ${\cal O}(p^4)$ is
incomplete because, at this stage, we do not have sufficient information to
constrain the full set of available parameters contributing at ${\cal O}(p^4)$.

   The EOMS scheme as well as the reformulated version of
the infrared renormalization \cite{Schindler:2003xv}, in principle, allow a
consistent analysis in a manifestly Lorentz-invariant framework even beyond the
one-loop level \cite{Schindler:2003je}.
   However, what seems to be equally important in the SU(3) sector is a consistent
consideration of the baryon decuplet \cite{Jenkins:1991ts}.
   Such an inclusion of spin $3/2$ into a manifestly Lorentz-invariant effective
field theory is a challenge, because it is a theory with constraints in order to
generate the correct number of degrees of freedom.

\ack
   J Gegelia acknowledges the support of the Alexander von Humboldt
Foundation and the Deutsche Forschungsgemeinschaft (SFB 443).

\appendix

\section{\label{integrals} Loop integrals}
   In this appendix we collect the dimensionally regularized loop integrals
needed in the calculation of the baryon masses in section \ref{O3calculation}.
   In what follows, $R$ is defined as
\begin{equation}
\label{R} R = \frac{2}{n-4}-[\ln(4\pi)+\Gamma'(1)+1]
\end{equation}
and we do not display terms of $O(n-4)$ and higher.
   From the set of purely mesonic integrals we only need
\begin{equation}
\label{Ipi} I_{M}= i\mu^{4-n}\int \frac{d^n k}{(2\pi)^n}\frac{1}{k^2-M^2+i0^+}
=\frac{M^2}{16\pi^2}\left[R+\ln\left(\frac{M^2}{\mu^2}\right)\right],
\end{equation}
   where $M$ is a meson mass at lowest order (see equation (\ref{Md2})).
   The $\widetilde{\rm MS}$-renormalized integral is obtained by simply
dropping the term proportional to $R$:
\begin{equation}
\label{Ipir} I_{M}^r=\frac{M^2}{8\pi^2}\ln\left(\frac{M}{\mu}\right).
\end{equation}
   The corresponding baryon integrals $I_B$ and $I_B^r$ are obtained from
equations (\ref{Ipi}) and (\ref{Ipir}) by the replacement $M\to M_0$.
   Note that, by choosing the scale $\mu=M_0$, the
$\widetilde{\rm MS}$-renormalized integral $I_B^r$ vanishes.
   Finally, the relevant integral containing both a meson and a baryon
   propagator reads
\begin{eqnarray}
\label{INpi} I_{BM}(-p,0)&=&i\mu^{4-n}\int\frac{d^n
k}{(2\pi)^n}\frac{1}{[(k-p)^2-M^2_0+i0^+]
[k^2-M^2+i0^+]}\nonumber\\
&=&\frac{1}{16\pi^2}\left[R-1+\ln\left(\frac{M_0^2}{\mu^2}\right)
+\frac{p^2-M^2_0+M^2}{p^2}\ln\left(\frac{M}{M_0}\right)\right.\nonumber\\
&&\left. +\frac{2M_0M}{p^2}F(\Omega)\right],
\end{eqnarray}
where
\begin{eqnarray*}
F(\Omega) &=& \left \{ \begin{array}{ll}
\sqrt{\Omega^2-1}\ln\left(-\Omega-\sqrt{\Omega^2-1}\right),&\Omega\leq -1,\\
\sqrt{1-\Omega^2}\arccos(-\Omega),&-1\leq\Omega\leq 1,\\
\sqrt{\Omega^2-1}\ln\left(\Omega+\sqrt{\Omega^2-1}\right)
-i\pi\sqrt{\Omega^2-1},&1\leq \Omega,
\end{array} \right.
\end{eqnarray*}
with
\begin{displaymath}
\Omega=\frac{p^2-M^2_0-M^2}{2M_0M}.
\end{displaymath}
   Again, the $\widetilde{\rm MS}$-renormalized integral is obtained by omitting
the $R$ term.

\section{\label{coefficients}Coefficients}
   The coefficients ${\cal F}_{BM}$ of equation (\ref{renlooppb}) are given by
\begin{equation}
\eqalign{
{\cal{F}}_{N \pi}&=-\frac{3}{4}\left[D^2+2DF+F^2\right],\\
{\cal{F}}_{N K}&=-\left[\frac{5}{6}D^2-DF+\frac{3}{2}F^2\right],\\
{\cal{F}}_{N \eta}&=
-\frac{1}{2}\left[\frac{1}{6}D^2-DF+\frac{3}{2}F^2\right],\\
{\cal{F}}_{\Sigma\pi}&=-\left[\frac{1}{3}D^2+2F^2\right],\\
{\cal{F}}_{\Sigma K}&=-\left[D^2+F^2\right], \\
{\cal{F}}_{\Sigma\eta}&=-\frac{1}{3}D^2,\\
{\cal{F}}_{\Lambda\pi}&=-D^2,\\
{\cal{F}}_{\Lambda K}&=-\left[\frac{1}{3}D^2+3F^2\right],\\
{\cal{F}}_{\Lambda\eta}&=-\frac{1}{3} D^2,\\
{\cal{F}}_{\Xi\pi}&=-\frac{3}{4}\left[D^2-2DF+F^2\right],\\
{\cal{F}}_{\Xi K}&=-\left[\frac{5}{6}D^2+DF+\frac{3}{2}F^2\right],\\
{\cal{F}}_{\Xi\eta}&=-\frac{1}{2}\left[\frac{1}{6} D^2+DF
+\frac{3}{2}F^2\right].}
\end{equation}

\end{document}